\documentclass[numbers]{elsarticle}
\usepackage{latexsym}
\usepackage[latin2]{inputenc}
\usepackage{longtable}
\usepackage{url}
\RequirePackage{lineno} 

\makeatletter
\def\ps@pprintTitle{%
 \let\@oddhead\@empty
 \let\@evenhead\@empty
 \def\@oddfoot{\centerline{\thepage}}%
 \let\@evenfoot\@oddfoot}
\makeatother

\begin{document}

\title{The Graph of Our Mind}

\author[p]{Balázs Szalkai}
\ead{szalkai@pitgroup.org}
\author[p]{Bálint Varga}
\ead{varga@pitgroup.org}
\author[p,u]{Vince Grolmusz\corref{cor1}}
\ead{grolmusz@pitgroup.org}
\cortext[cor1]{Corresponding author}
\address[p]{PIT Bioinformatics Group, Eötvös University, H-1117 Budapest, Hungary}
\address[u]{Uratim Ltd., H-1118 Budapest, Hungary}


\date{}

\begin{abstract}
Graph theory in the last two decades penetrated sociology, molecular biology, genetics, chemistry, computer engineering, and numerous other fields of science. One of the more recent areas of its applications is the study of the connections of the human brain. By the development of diffusion magnetic resonance imaging (diffusion MRI), it is possible today to map the connections between the 1-1.5 cm$^2$ regions of the gray matter of the human brain. These connections can be viewed as a graph. We have computed 1015-vertex graphs with thousands of edges for hundreds of human brains from one of the highest quality data source of the Human Connectome Project. Here we analyze the male and female braingraphs graph-theoretically, and show statistically significant differences in numerous parameters between the sexes: the female braingraphs are better expanders, have more edges, larger bipartition widths, and larger vertex cover than the braingraphs of the male subjects. We apply the data of 426 subjects, and demonstrate the statistically significant (corrected) differences in 116 graph parameters between the sexes.
\end{abstract}

\maketitle

\section*{Introduction} 

It is an old dream to describe the neuronal-level braingraph (or connectome) of different organisms, where the vertices correspond to the neurons and two neurons are connected by an edge if there is a connection between them. The connectome of the roundworm {\em Caenorhabditis elegans} with 302 neurons was mapped 30 years ago \citep{White1986}, but larger braingraphs, especially the complete fruitfly {\em Drosophila melanogaster} braingraph (the "flybrain") with approximately 100,000 neurons remained unmapped in its entirety, despite using enormous resources and efforts worldwide. Mapping the connections in the human brain on the neuronal level is completely hopeless today, mostly because there are, on the average, 86 billion neurons in the human brain \citep{Azevedo2009}.
Constructing human braingraphs (or "connectomes"), where the vertices are not single neurons, but much larger areas of the gray matter of the brain (called Regions of Interest, ROIs), is possible, and it is the subject of a very intensive research work today. Two vertices, corresponding to the ROIs, are connected by an edge if a diffusion-MRI based workflow finds neuronal connections between them. In the process of the Human Connectome Project \citep{McNab2013}, an enormous amount of data and numerous tools were created related to the mapping of the human brain, and the resulting data were deposited in publicly available databases of dozens of terabytes. 

Our focus in this work is the graph theoretical analysis of the connections of the brain; consequently, we just sketch the process of the construction of this graph here.

The human brain tissue, roughly, has two distinct parts: the white matter and the gray matter. The gray matter, by some simplifications, consists of the cell-bodies (or {\em somas}) of the neurons, and the white matter from the fibers of {\em axons} (long projections from the somas), insulated by lipid-like myelin sheaths. The cortex of the brain, and also some sub-cortical areas, contain gray matter, and most of the inner parts of the brain contain white matter. Again with some simplifications, the connections between the somas of the neurons, the axons, run in the white matter, except the very short axons running entirely in the gray matter.

Diffusion magnetic resonance imaging (MRI) is, again roughly speaking, capable of measuring the direction of the diffusion of the water molecules in living tissues, without any contrast agent. The gray matter of the brain consists of the cell bodies (somas) of the neurons, consequently, there is not any distinguished direction of the diffusion of the water molecules in the somas: in each direction the molecules can move freely. In the white matter, however, the neuronal fibers consisted of long axons, so the water molecules move more easily and more probably in the direction of the axons than perpendicularly, through the cell membrane, bordering the axons. Therefore, in each point of a given axon in the white matter, the diffusion of the water molecules is larger in directions parallel to the axons and smaller in other directions. 

This way one can distinguish the white matter and the gray matter of the brain (this step is called partitioning). Moreover, by following or tracking the directions of the stronger diffusion, it is possible to map the orbits of the neuronal fibers in the white matter (this step is called tractography). Certainly, when the fiber tracts are crossed, it is not easy to follow the correct directions of the axons. 

After the tractography is performed, one gets an image, similar to Figure 1. Most of the fibers start and end on the surface --- the cortex --- of the brain.

We are interested in the connections between the gray matter areas, mostly of the cortical areas, and we ignore the exact orbits of the neuronal fibers in the white matter. That is, it is not interesting for us where the ``wires'' run, just the fact of the connections between the separate areas of the gray matter. Naturally, the length or the number of neuronal fibers, connecting the gray matter ROIs, can be included in the graph as different weight functions on the edges.

Consequently, we define the graph as follows: the vertices are the small anatomical areas of the gray matter (ROIs), and two ROIs are connected by an edge if in the tractography phase, at least, one fiber is tracked between these two ROIs. We are considering five different resolutions of ROIs, and also five different weight functions, computed from the properties of the fibers, connecting the ROIs.

\begin{figure} [h!]
	\centering
	\includegraphics[width=5.2in]{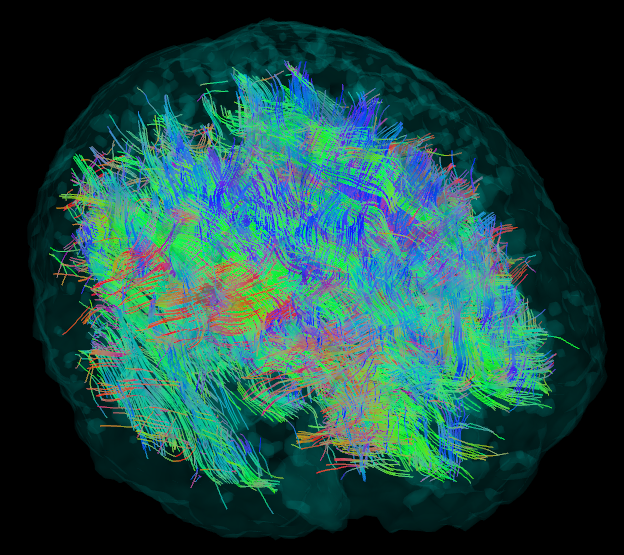}
	\caption{The result of the tractography phase. Note the fibers starting and ending on the outer surface, the cortex of the brain, which consists of gray matter. The fibers are tracked only in the anisotropic white matter.}
\end{figure}

\subsection*{Previous work}

Numerous publications cover the connectome \citep{Hagmann2012,Craddock2013a} of healthy humans \citep{Ball2014,Bargmann2012,Batalle2013,Graham2014} and also the connectomes of the healthy and the diseased brain \citep{Agosta2014,AlexanderBloch2014,Baker2014,Besson2014a,Bonilha2014}. Usually, these works analyze only 80-100 vertex graphs on the whole brain, and they are using concepts that originate from the network science, developed for large graphs of millions of vertices, found, e.g., in the graph of the World Wide Web.

Here we present another approach: We are analyzing larger graphs, up to 1015 vertices, and our algorithms are originated from graph theory and not from network science. In other words, we are also computing graph parameters that are quite hopeless to compute for graphs with millions of vertices. 

In our previous work, we have made comparisons between the braingraphs of numerous subjects with several focuses:

\begin{itemize}
	\item[(i)] We have constructed the Budapest Reference Connectome Server \url{http://connectome.pitgroup.org}, which generates the common edges of up to 477 graphs of 1015 vertices, according to selectable parameters \citep{Szalkai2015a,Szalkai2016}. The Budapest Reference Connectome Server, apart from the common-edge demonstration, is also a good tool for the instant visualization of the braingraph.
	
	\item[(ii)] We have compared the diversity of the edges in distinct cerebral areas in 392 individual brains in \citep{Kerepesi2015c};
	
	\item[(iii)] Based on a feature of the Budapest Reference Connectome Server, we have found a probable connection between the consensus connectomes and the axon--development of the human brain
	 \citep{Kerepesi2015b,Kerepesi2016,Szalkai2016d,Szalkai2016e}. 
	 
	 \item[(iv)] We have described the most frequent small subgraphs of the human braingaph in \cite{Fellner2017}. In \cite{Fellner2019} we have listed the most frequent complete subgraphs of the human connectome. In \cite{Fellner2018,Fellner2019a} we have introduced the method of the Frequent Network Neighborhood Mapping, and applied it for the neighbors of the hippocampus, one of the most important small funcional entity of the brain. 
	
	\item[(v)] We have compared women's and men's connectomes in 96 subjects in \citep{Szalkai2015}, and found that the braingraphs of females have numerous, statistically significant advantages in graph theoretical properties that are characteristic of the better connections. We have found 13 parameters, in which the difference remained significant after the very strict Holm-Bonferroni statistical correction \citep{Holm1979}.
	
\end{itemize}

In the present work, we have found 116 graph parameters (vs. the 13 parameters in \citep{Szalkai2015}), which differ significantly between the sexes, after the Holm-Bonferroni correction.

\section*{Results}

In the present work we are considering a 426-subject dataset from the Human Connectome Project public release \citep{McNab2013}. For each subject, we compute braingraphs with five different vertex-numbers: 83, 129, 234, 463 and 1015. The vertices correspond to anatomical areas of the gray matter in different resolutions. 

The goal is to assign the same named vertex to the same anatomical region, in the case of all subjects. Naturally, the size and the shape of the brain of all subjects differ; therefore it is a non-trivial task to assign the same nodes (or ROIs) to the same anatomical regions for all subjects. This is called the ``registration problem'', and we have applied the solution from the FreeSurfer suite of programs \citep{Fischl2012} that mapped the individual brains to the  Desikan-Kiliany brain atlas \citep{Desikan2006}. Roughly, the registration method applies homeomorphisms in order to correspond the major cortical patterns of {\em sulci} and {\em gyri} between individual cortices. 

We were using five different resolutions in 83, 129, 234, 463 and 1015 vertices,  because for smaller values the graph structure is poorer, and for the higher resolutions there is a possibility of the registration errors, due to the potentially too small areas corresponded to the vertices. Therefore, we have computed and analyzed the graph properties for all of these five resolutions, instead of deciding arbitrarily that one of the resolutions are the best for our goals.

For every graph, we have considered five different edge weights. Four of these describe some quantities, related to the neuronal fibers, defining the edge in question. More exactly, the weight functions are:

\begin{itemize}
	\item \texttt{Unweighted}: Each edge has the same weight 1;
	\item \texttt{FiberN}: The number of fibers discovered in the tractography step between the nodes, corresponded to ROIs;
	\item \texttt{FAMean}: The average of the fractional anisotropies \citep{Basser2011} of the neuronal fibers, connecting the endpoints of the edge;
	\item \texttt{FiberLengthMean}: The average fiber-lengths between the endpoints of the edge.
	\item \texttt{FiberNDivLength}: The number of fiber tracts connecting the end-nodes, divided by the mean length of those fibers. 
\end{itemize}
 
 The last weight function, \texttt{FiberNDivLength}, describes a conductance-like quantity in a very simple electrical model: the resistances is proportional to the average fiber length and inversely proportional to the number of wires connecting the endpoints. Similarly, it is also describing a reliability measure of the edge: longer fibers are less reliable due to tractography errors, but multiple fibers between the same ROIs are increasing the reliability.

 Other authors have considered the number of edges (weighted or unweighted), running between pre-defined areas of the brain. One of the main focuses of these works was the ratio between the edges, running between the two hemispheres of the brain divided by the number of edges running within each hemisphere \citep{Sex2011,Ingalhalikar2014b}. The authors of \citep{Ingalhalikar2014b} considered 95-node graphs, computed from 949 subjects of a publicly unavailable dataset,  and found that, relatively, males have more intra-hemispheric edges while females have more inter-hemispheric edges. 
 
 We were interested --- instead of simple edge-counting between pre-defined vertex-sets --- in computing much more elaborate graph-theoretic parameters of the braingraphs. 
 
 More exactly, we have computed the following parameters, for each graph, similarly as in \citep{Szalkai2015,Szalkai2015c}:

 \begin{itemize}
 	\item Number of edges (\texttt{Sum}). The weighted version of this number is the sum of the weights of the edges in the graph.

 	\item Normalized largest eigenvalue (\texttt{AdjLMaxDivD}): The largest eigenvalue of the generalized adjacency matrix, divided by the average degree of the graph. The adjacency matrix of an $n$-vertex graph is an $n\times n$ matrix, where $a_{ij}$ is 1 if $\{v_i,v_j\}$ is an edge, and 0 otherwise. The generalized adjacency matrix contains the weight of edge $\{v_i,v_j\}$ in $a_{ij}$. The division by the average degree of the vertices is important since the largest eigenvalue is bounded by the average- and maximum degrees \citep{Lovasz2007}, so a dense graph has a big $\lambda_{max}$ largest eigenvalue because of the larger average degree. Since the vertex numbers are fixed, the average degree is already defined by the sum of weights for each graph.

 	\item Eigengap of the transition matrix (\texttt{PGEigengap}): The transition matrix $P_G$ is defined by dividing the rows of the generalized adjacency matrix by the generalized degree of the node, where the generalized degree is the sum of the weights of the edges, incident to the vertex. A random walk on a graph can be characterized by the probabilities, for each $i$ and $j$, of moving from vertex $v_i$ to vertex $v_j$. These probabilities are the elements of transition matrix $P_G$, with all the row-sums equal to 1. The eigengap of a matrix is the difference between the largest and the second largest eigenvalue of $P_G$, and it is characteristic of the expander property of the graph: the larger the gap, the better expander is the graph (see \citep{Hoory2006}).

 	\item Hoffman's bound (\texttt{HoffmanBound}): If  $\lambda_{max}$ and $\lambda_{min}$ denote the largest and smallest eigenvalues of the adjacency matrix, then Hoffman's bound is defined as $$1 + \frac{\lambda_{max}}{|\lambda_{min}|}.$$ This quantity is a lower estimation for the chromatic number of the graph.

 	\item Logarithm of the number of the spanning forests (\texttt{LogAbsSpanningForestN}): The quantity of the spanning trees in a connected graph can be computed from the spectrum of its Laplacian \citep{Kirchoff1847,Chung1997}. Graphs with more edges usually have more spanning trees, since the addition of an edge does not decrease the number of the spanning trees. For non-connected graphs, the number of spanning forests is the product of the numbers of the spanning trees of their components. The quantity \texttt{LogAbsSpanningForestN} is defined to be the logarithm of the number of spanning forests in the unweighted case. For other weight functions, if we define the weight of a tree by the product of the weights of its edges, then \texttt{LogAbsSpanningForestN} equals to the sum of the logarithms of the weights of the spanning trees in the forests.

 	\item Balanced minimum cut, divided by the number of edges (\texttt{MinCutBalDivSum}): If the nodes of a graph are partitioned into two classes, then a {\em cut} is the set of the edges running between these two classes. When we are looking for a {\em minimum cut} in a graph, most frequently one of the classes is small (say it contains just one vertex) and the other all the remaining vertices. Therefore, the most interesting case is when the sizes of the two classes of the partitions differ by at most one. Finding such a partition with the smallest cut is the "balanced minimum cut" or the "minimal bisection width" problem. This quantity, in a certain sense, describes the "bottleneck" of the graph, and it is an important characteristic of the interconnection networks (like the butterfly, the cube connected cycles, or the De Bruijn network, \citep{Tarjan1983a}) in computer engineering.  For the whole brain graph, one may expect that the minimum cut corresponds to the partition to the two hemispheres, which was found when we analyzed the results. Consequently, this quantity is interesting {\em within} the hemispheres, when only the nodes of the right- or the left hemisphere is partitioned into two classes of equal size. Computing the balanced minimum cut is NP-hard \citep{Garey1976}, but its computation for the input-sizes of this study is possible with contemporary integer programming software. If we double every edge in a graph (allowing two edges between two vertices) then the minimum balanced cut will also be doubled. So, it is natural to expect that graphs with more edges may have larger minimum balanced cut just because the more edges present. However, if we {\em norm (i.e., divide by) } the balanced minimum cut with the number of the edges in the graph examined, then this effect can be factored out: for example, in the doubled-edge graph the balanced minimum cut is also doubled, but when its size is divided by the doubled edge number, the normed value will be the same as in the original graph. So, when \texttt{MinCutBalDivSum} is considered, the effects of the edge-numbers are factored out.

 	\item Minimum cost spanning tree (\texttt{MinSpanningForest}), computed with Kruskal's algorithm \citep{Lawler1976}. 
 	
 	\item Minimum weighted vertex cover (\texttt{MinVertexCover}): We need to assign to each vertex a non-negative weight satisfying that for each edge, the sum of the weights of its two endpoints is at least 1. This is the relaxation of the NP-hard vertex-cover problem \citep{Hochbaum1982}, since here we allow fractional weights, too. The sum of all vertex-weights with this constraint can be minimized in polynomial time by linear programming.
 	
 	\item Minimum vertex cover (\texttt{MinVertexCoverBinary}): Same as the quantity above, but the weights need to be 0 or 1. Alternatively, this number gives the size of the smallest vertex-set such that each edge is connected to at least one of the vertices in the set. This graph parameter is NP-hard, and we computed it only for the unweighted case by an integer programming (IP) solver SCIP \url{http://scip.zib.de} \citep{Achterberg2008, Achterberg2009}. 
 	
 	\item Maximum matching (\texttt{MaxMatching}): A graph matching is a set of edges without common vertices. A maximum matching contains the largest number of edges. A maximum matching in a weighted graph is the matching with the maximum sum of weights taken on its edges.
 	
 	\item Maximum fractional matching (\texttt{MaxFracMatching}): is the linear-programming relaxation of the maximum matching problem. In the unweighted case,  non-negative values $x(e)$ are searched for each edge $e$ in the graph, satisfying that for each vertex $v$ in the graph, the sum of $x(e)$-s for the edges that are incident to $v$ is at most 1. The maximum of the sums of $\sum_ex(e)$ is the maximum fractional matching for a graph. For the weighted version with weight function $w$, $\sum_ex(e)w(e)$ needs to be maximized.

 \end{itemize}
 
 The above parameters were computed for all five resolutions and the left and the right hemispheres and also for the whole connectome, with all five weight functions (with the following exceptions: \texttt{MinVertexCoverBinary} was computed only for the unweighted case, and the \texttt{MinSpanningTree} was not computed for the unweighted case). 
 
 The results, for each subject, each resolution and each weight function are detailed in a large Excel table, downloadable from the site \url{http://uratim.com/bigtableB.zip}. 
 
 \subsection*{The syntactics of the results:} 

Each parameter-name in the table at \url{http://uratim.com/bigtableB.zip} and elsewhere in this work contains two separating ``$\_$'' symbols that define three parts of the name. The first part describes the hemisphere or the whole connectome with the words Left, Right or All. The second part describes the parameter computed, and the third part the weight function used. For example, \texttt{All\_AdjLMaxDivD\_FiberNDivLength} means that the normalized largest eigenvalue \texttt{AdjLMaxDivD} was computed for the whole brain, with the  \texttt{FiberNDivLength} weight function (see above).

In the Table \url{http://uratim.com/bigtableB.zip}, the first column, round index is used in the statistical analysis. Second column, ``id'', is the anonymized subject ID of the Human Connectome Project's 500-subject public release. Column 3 gives the sex of the subject, 0: female, 1: male. Fourth column gives the age-groups 0: 22-25 years; 1: 26-30 years; 2: 31-35 years; 3: 35+ years.  Column 5 gives the number of vertices of the graph analyzed. 

\section*{Discussion} The data that we used from the public release of the Human Connectome project, contains diffusion MRI recordings from healthy male and female subjects of age 22 through 35. Therefore, if we want to find correlations of the graph theoretical characteristics of the connectomes with some biological properties, we may easily use either the sex or the age of the subjects. 

Our main finding now, on a large data set, validates our earlier results that was made on a much smaller data set in \citep{Szalkai2015}: in numerous graph theoretical parameters, women's connectomes show statistically significant advantages against the men's respective parameters. The parameters in question are related to ``better connectivity'' in several aspects. 

In the supporting material, we are enclosing several large tables with the results. In Table 1, the results of statistical analysis are detailed: the parameters with the bold last column are {\em all} significantly differ between the female and the male connectomes: the vast majority is ``better'' for the females. If the last column is not bold, but the fifth column is typeset in italic then those parameters, one-by-one, significantly differ between the sexes, but it is unlikely that all of them differ significantly (type II statistical errors are possible).

For example, as it is seen in Table 1, differences in the \texttt{PGEigengap} values show the better expander property in the braingraph of the females, in both hemispheres. The differences in the \texttt{Sum} quantity shows that in both hemispheres, women have more edges than men, and this statement remains true for weighted edges with most weight functions. Very strong statistical evidence show the difference and the women's advantage in the edge-number normalized balanced minimum cut in the left hemisphere. Matching numbers (both fractional and integer) are also significantly larger in the case of females.

Seemingly, in the left hemisphere the women's advantage is stronger in several parameters: the first several rows of Table 1 contains mostly ``Left'' or ``All'' prefixes in the second column.

In very few cases men have better parameters: e.g., in resolution 83, \texttt{All\_MinSpanningForest\_FiberLengthMean} is significantly larger for men than for women. Similarly, another parameter, weighted by \texttt{FiberLengthMean}, the \texttt{All\_MinSpanningForest\_FiberLengthMean} in 234-resolution is also larger for males. We believe that the larger brain size with the \texttt{FiberLengthMean} weighting compensates the fewer connections of the males in these cases.

In the supporting material, we are also enclosing Tables 2, 3, 4, 5 and 6 that give the detailed averaged results for each resolution for each graph parameter with ANOVA statistical analysis. The subject-level data are also available at \url{http://uratim.com/bigtableB.zip}.

\section*{Materials and Methods}

We have used the Connectome Mapper Toolkit \citep{Daducci2012} \url{http://cmtk.org} for brain tissue segmentation into gray and white matter, partitioning the brain into anatomical regions, for tractography (tracking the axonal fibers in the white matter) and for the construction of the graphs from the fibers identified in the tractography phase of the workflow. The partitioning was based on the FreeSurfer suite of programs \citep{Fischl2012}, according to the Desikan-Killiany brain anatomy atlas \citep{Desikan2006}. The tractography used the MRtrix processing tool \citep{Tournier2012} with randomized seeding and with the deterministic streamline method. 
 
The graphs were constructed using the results of the tractography step: two nodes, corresponding to ROIs, were connected if there existed, at least one, fiber connecting them. Loops were deleted from the graph.

Graph parameters were computed by the integer programming (IP) solver SCIP \url{http://scip.zib.de}, \citep{Achterberg2008, Achterberg2009}, and by some in-house scripts.

The unprocessed and pre-processed MRI data is available at the Human Connectome Project's website: \url{http://www.humanconnectome.org/documentation/S500}  \citep{McNab2013}. The assembled graphs that we analyzed in the present work can be downloaded at the site \url{http://braingraph.org/download-pit-group-connectomes/}. The individual graph results are detailed in a large Excel table at the site \url{http://uratim.com/bigtableB.zip}

\subsection*{Statistical analysis}

Our statistical null-hypothesis \citep{Hoel1984} was that the graph parameters do not differ between males and females. For dealing with both type I and type II statistical errors, we have partitioned the subjects into classes quasi-randomly: subjects with IDs with even digit-sums went to group 0, and those with odd digit sums went to group 1 (c.f. the first column of \url{http://uratim.com/bigtableB.zip}). 

We applied group 0 for a base set, for making hypotheses, and group 1 as a holdout set, for testing those hypotheses. The hypotheses on group 0 were filtered by ``Analysis of variance'' (ANOVA) \citep{Wonnacott1972}: only the hypotheses with p-value of less than 1\% were selected for the testing in the holdout set. Next, the selected hypotheses were tested on group 1, with the rather strict Holm-Bonferroni correction method \citep{Holm1979}. The significance level in the Holm-Bonferroni correction was set to 5\%.

\subsection*{Handling possible artifacts}

While we have applied the same computational workflow for the data of the both sexes, it is still possible that some non-sex specific artifact caused the significant differences in the graph parameters between men and women subjects. One possible cause may be the statistical difference between the size of the brain of the sexes \citep{Witelson2006}. In the tractography step, it may happen that the longer neural fibers of the males cannot be tracked so reliably as the shorter fibers of the females. To close out this possible error, we have selected 36 small-brain males and 36 large-brain females such that all the females have larger brains than all the males in the data set \citep{Szalkai2015c}. Next, we have computed the graph theoretical parameters as in the present work. Two main findings of ours were: (i) the small-brain men did not have the advantages identified in the set of the women in the present study; (ii) in several parameters, mostly with the weight function \texttt{FAMean}, women still have the statistically significant advantages identified in the present study. 

We find this result decisive that the graph-theoretical differences in the connectomes are due to sex differences and not size differences. 


\section*{Data accessibility:} The unprocessed and pre-processed MRI data is available at the Human Connectome Project's website: \url{http://www.humanconnectome.org/documentation/S500}  \citep{McNab2013}. The assembled graphs that we analyzed in the present work can be downloaded at the site \url{http://braingraph.org/download-pit-group-connectomes/}. The individual graph results are detailed in a large Excel table at the site \url{http://uratim.com/bigtableB.zip}



\section*{Funding:}  VG and BV were partially funded by the VEKOP-2.3.2-16-2017-00014 program, supported by the European Union and the State of Hungary, co-financed by the European Regional Development Fund, and by the 
European Union, co-financed by the European Social Fund (EFOP-3.6.3-VEKOP-16-2017-00002). VG was partially funded by the NKFI-126472 and NKFI-127909
 grants of the National Research, Development and Innovation Office of Hungary. 

\section*{Acknowledgments:}
Data were provided in part by the Human Connectome Project, WU-Minn Consortium (Principal Investigators: David Van Essen and Kamil Ugurbil; 1U54MH091657) funded by the 16 NIH Institutes and Centers that support the NIH Blueprint for Neuroscience Research; and by the McDonnell Center for Systems Neuroscience at Washington University.




\end{document}